\documentclass[
]{ceurart}

\usepackage{listings}
\usepackage{comment}
\usepackage{enumitem}
\begin{document}

\copyrightyear{2026}
\copyrightclause{Copyright for this paper by its authors.
  Use permitted under Creative Commons License Attribution 4.0
  International (CC BY 4.0).}

\conference{International Symposium on Artificial Intelligence and the Transformation of Higher Education Symposium (AITHE),
  July 01--03, 2026, Swansea, Wales, UK}

\title{Structuring Transparency: Developing Domain-Specific Generative AI Declaration Frameworks in Higher Education}


\author[1]{Nicholas Micallef}[%
orcid=0000-0002-8668-2019,
email=nicholas.micallef@swansea.ac.uk,
url=https://www.swansea.ac.uk/staff/nicholas.micallef
]
\cormark[1]

\author[1,2]{Olga Petrovska}[%
orcid=0000-0003-1170-8816,
email=olga.petrovska@swansea.ac.uk, 
url=https://opetrovska.github.io,
]
\address[1]{Swansea University, Department of Computer Science, Computational Foundry, 
Bay Campus, Swansea, Wales, United Kingdom}
\address[2]{Institute of Coding in Wales,
CoSMOS, Margam Building, Swansea University, Singleton Park, Swansea, Wales, United Kingdom}

\cortext[1]{Corresponding author.}

\begin{abstract}
As Generative AI (GenAI) disrupts higher education, institutions increasingly require students to declare AI use. However, generic, binary declarations (e.g., ``I used GenAI") fail to capture the nuanced application of these tools in different academic tasks. Establishing transparency is key to protecting academic integrity, promoting AI literacy, and shifting the focus from policing to professional practice. 
In response, this paper contributes a design artefact and an accompanying position: a framework of two task-specific declaration structures, one for writing-focused activities and one for coding assessments, developed for a Computer Science department on the basis of an existing taxonomy of GenAI usage~\cite{buovier2026generative}, together with an argument that task-specific disclosure is needed to move beyond binary declarations.
By categorising AI usage across specific cognitive and developmental stages, such as structural planning vs. Textual Content Generation, or code improvement vs. code generation, the framework encourages students to reflect on their own learning process and clarifies the boundary between acceptable assistance and academic misconduct. We propose this domain-specific approach as a foundation for fostering more honest assessment in Computer Science and other disciplines, aiming to better prepare students for professional environments where documenting GenAI workflows might be an essential job requirement.
\end{abstract}


\maketitle

\section{Introduction}
Generative AI (GenAI) tools such as ChatGPT, Claude, and Copilot are rapidly reshaping how students learn, write, and program in higher education~\cite{ravselj2025higher}. In Computer Science (CS) in particular, students now routinely encounter GenAI both as a topic of study and as a set of tools that can be integrated into their everyday workflows~\cite{prather2024widening}. This, coupled with the ready availability of numerous GenAI tools, places increasing pressure on universities to prepare graduates for professional environments where GenAI use is becoming normalised, while also upholding academic integrity of assessments~\cite{francis2025generative}.

Many universities have responded by introducing AI policies and mandatory student declarations of GenAI use~\cite{moorhouse2023generative,wilson2025ethical}. However, these declarations are often generic and binary - for example, a single checkbox stating whether GenAI was used in an assignment~\cite{gonsalves2025addressing, maguire2025themes}. Such approaches fail to capture the diversity of ways in which GenAI can be involved in academic work, from brainstorming and clarifying requirements to generating code, drafting text, or polishing language~\cite{buovier2026generative,perkins2024artificial}. 
This matters for assessments: when a declaration cannot distinguish brainstorming from ghost-writing, it simultaneously fails as an integrity safeguard, as a reflective prompt for students, and as a meaningful source of information for educators adapting their practice.
Generic and binary declarations provide limited insight into how tools are actually being used and offer little pedagogical value to students or staff~\cite{gonsalves2025addressing}.

At the same time, emerging research on GenAI in computing education has begun to catalogue the different activities GenAI can support, identifying aspects such as generating artefacts, interpreting information, refining work, and obtaining feedback~\cite{buovier2026generative,hou2024effects}. These taxonomies offer a useful analytic lens, but are not directly integrated into assessment workflows. There remains a gap between high-level conceptualisations of GenAI use and the practical mechanisms through which students disclose and reflect on their own use of these tools~\cite{gonsalves2025addressing,maguire2025themes}.

This paper addresses the identified gap by presenting a domain-specific framework for structuring GenAI declarations in CS assessments. Building on an existing taxonomy of GenAI usage~\cite{buovier2026generative}, we develop two tailored declaration structures: one for writing-focused activities (such as professional or literature-review style reports) and another for programming-based assessments. Each structure breaks GenAI use into activity types and asks students to indicate both whether and to what extent they used GenAI, alongside short explanations and example prompts.

This paper is framed as a design contribution, offering a structured artefact for practice, a position on how task-specific disclosure can support more honest assessment, and an empirical agenda for its future evaluation.
While the framework is developed within a CS department, we suggest that the underlying approach is adaptable to other disciplines facing similar challenges amid the rise of GenAI.

\section{Background and Related Work}
In this section, we position our framework within four related strands of literature: the rapid adoption of GenAI by students in higher education, institutional policy and declaration responses, the concerns specific to computing education, and existing frameworks for structuring GenAI use in assessment.
 
\subsection{GenAI adoption in higher education}
Since the public release of ChatGPT in late 2022, GenAI tools have been adopted by students at an unprecedented pace. In the most comprehensive global study to date, involving 23,218 higher education students across 109 countries, Ravšelj et al.~\cite{ravselj2025higher} report that students predominantly use GenAI for brainstorming, summarising texts, and finding research articles, with smaller cohorts also using it for professional and creative writing. This pattern is important for our framework: students are not simply engaging with GenAI for a single task type, but distribute their use across planning, comprehension, and drafting stages of academic work. The same study finds that although students are broadly positive about ChatGPT as a learning tool, they also express concern about its potential to facilitate cheating and plagiarism~\cite{ravselj2025higher}, foreshadowing the policy tensions that institutions have subsequently faced.
 
These tensions are central to recent higher-education research, which frames GenAI as a double-edged phenomenon combining transformative pedagogical potential with genuine challenges to traditional assessment paradigms~\cite{francis2025generative}. Wilson~\cite{wilson2025ethical} examines the ethical foundations of institutional advice to students on GenAI, observing that much early guidance was reactive and grounded in prohibition rather than pedagogy. Responding to this, Symeou et al.~\cite{symeou2025development} develop evidence-based, consensus-built guidelines for integrating GenAI into university education, arguing that multidisciplinary collaboration is required to produce guidance that is both pedagogically coherent and practically enforceable.
 
\subsection{Institutional policies and AI use declarations}
A widely cited analysis by Moorhouse et al.~\cite{moorhouse2023generative} reviewed 50 top-ranked global higher education institutions and found that 57\% had issued guidelines on GenAI use, primarily centring around two requirements: acknowledgement of AI use and citation of AI-generated content. While this represents a sensible first step, subsequent empirical work has highlighted the limitations of these approaches in practice. Gonsalves~\cite{gonsalves2025addressing} reports a study at King's Business School in which a coursework coversheet included a mandatory AI use declaration; yet 74\% of students did not complete it, with interview data attributing non-compliance to fear of academic consequences, ambiguous guidelines, inconsistent enforcement, and peer influence. Students reported perceiving the declaration similar to \emph{admitting to plagiarism} rather than a neutral act of transparency, illustrating how coarse declaration mechanisms can inadvertently undermine the honesty they are intended to develop~\cite{gonsalves2025addressing}.
 
Maguire et al.~\cite{maguire2025themes}, working in a UK cyber-security context, analysed the themes that emerge when students do declare their GenAI use. They found that declarations tend to cluster into a small set of patterns, with many students describing similar surface-level uses (such as editing and proofreading) and few articulating how GenAI shaped their reasoning or design decisions. Their work reinforces the observation that binary declarations, or free-text boxes without scaffolding, produce neither the granularity nor the diversity of information that educators need to understand GenAI workflows~\cite{maguire2025themes}.
 
\subsection{GenAI in computing education}
Computing education research has been particularly active in documenting how GenAI reshapes student practice, and in proposing conceptual structures to describe it. In a report focused on GenAI interventions in post-introductory computing courses, Bouvier et al.~\cite{buovier2026generative} propose a taxonomy of nine GenAI-mediated activities.
The taxonomy is designed to support comparative analysis of interventions across sub-domains of computing; importantly for our purposes, it is an analytic instrument rather than a disclosure instrument. Complementary work by Hou et al.~\cite{hou2024effects} characterises the ``context of usage'' of GenAI in computing education, identifying four common student-facing use cases: understanding course concepts, debugging code, identifying corner cases, and writing code.
 
Empirical studies of how novices actually use GenAI tools further motivate the need to differentiate modes of use. In an eye-tracking and think-aloud study of novice programmers, Prather et al.~\cite{prather2024widening} observe a ``widening gap'' in outcomes: students with stronger metacognitive skills use GenAI to accelerate work they could already conceive, while struggling students experience new metacognitive difficulties, including interruption, being misled by suggestions, and an illusion of competence. The implication is that a single label such as ``AI use'' obscures qualitatively different modes of engagement. 
 
\subsection{Frameworks for structuring GenAI use in assessment}
The most prominent framework for structuring GenAI use in assessment is the AI Assessment Scale (AIAS) developed by Perkins et al.~\cite{perkins2024artificial}. The AIAS defines five ordered levels of permitted GenAI use, ranging from ``No AI'' to ``AI Exploration'', and is intended as a communication mechanism between educators and students about what is allowed in a given task. Since its introduction, the AIAS has been adopted by institutions across the world and translated into over thirty languages, and its authors have subsequently revised it in light of growing technological capability and pedagogical experience~\cite{perkins2024artificial}.
 
The AIAS and our framework address complementary problems. The AIAS operates primarily at the level of the \emph{educator's design decision}: what level of GenAI assistance is permitted in a given assessment; whereas our framework operates at the level of the \emph{student's declaration}: how GenAI was used within a task whose permitted scope has already been defined. The computing-specific taxonomies discussed above~\cite{buovier2026generative,hou2024effects} sit at a third, more analytic level: they describe what kinds of activity occur, without either prescribing what is allowed or providing a structure for disclosure. Our contribution sits deliberately at the intersection of these concerns, translating an activity-level taxonomy into task-specific disclosure instruments that can operate within any AIAS level where some GenAI use is permitted.
 

\section{Context: Computer Science Assessment Needs}
Computer Science education spans multiple programmes and modes of study. 
Within a single department, cohorts may include traditional school-leavers, international students, career-changers, and experienced practitioners studying through degree apprenticeship schemes. This results in a  diversity of prior programming experience, professional practice, and varied 
levels of engagement with GenAI.  

Assessment practices in CS must, therefore, accommodate a wide variety of learner profiles and expectations. Students encounter diverse task types, ranging from technical coding assignments and software projects to written reports on professional, ethical, or socio-technical issues; correspondingly, they engage with GenAI in varied ways. In this context, simplistic, binary declarations of AI use obscure important differences between legitimate, context-appropriate assistance and uses that undermine the intended learning outcomes. This motivates the need for domain- and task-specific approaches to structuring transparency around GenAI use in CS assessment.

This heterogeneity motivates a task-specific rather than a module- or programme-level approach to declarations. A single institutional form cannot simultaneously serve a degree apprentice writing a professional reflection, a student debugging a legacy codebase, and a novice learning Python syntax, as each involves legitimate but qualitatively different GenAI use profiles. It also reflects the heterogeneity of professional practice itself, where some employers mandate GenAI use, others prohibit it for intellectual-property or security reasons, and others require detailed documentation of prompts and outputs.

\section{Domain-Specific GenAI Declaration Framework}
In this section, we describe the domain-specific GenAI declaration framework developed for the CS context. We first outline the general design principles, then present the writing-focused and coding-focused structures derived from the underlying taxonomy of GenAI usage.
\subsection{Framework principles}
Our framework builds on the ``taxonomy of GenAI use in activities'' proposed by Bouvier et al., which characterises GenAI-mediated work in terms of aspects such as \emph{Generate, Interpret, Evaluate, Get Feedback, Refine, Brainstorm, Design, Simulate,} and \emph{Reflect}~\cite{buovier2026generative}. This taxonomy was developed from a systematic review of GenAI interventions in post‑introductory computing education and is intended to support comparison of GenAI use across subject areas~\cite{buovier2026generative}. We extend this work by translating these abstract aspects into concrete, assessment-facing categories that students and instructors can use when declaring GenAI use in specific coursework tasks.

The framework is guided by five principles that collectively shift declarations from compliance artefacts to reflective instruments:
\begin{itemize}[leftmargin=*,nosep]
\item \textbf{\emph{Granularity over a binary yes/no}}: Rather than a single ``Did you use GenAI?'' question, we distinguish between multiple activity types within each assessment, acknowledging that using GenAI to brainstorm, refine, or generate artefacts are qualitatively different practices.
\item \textbf{\emph{Task specificity}}: We provide separate declaration structures for writing-focused tasks and coding tasks, reflecting the different workflows and GenAI practices in each assessment type.
\item \textbf{\emph{Intensity as well as presence}}: For each activity type, students indicate whether they used GenAI and to which extent of  (\emph{Minor, Moderate, Extensive}), loosely reflecting the varying intensity with which aspects such as \emph{Generate} or \emph{Refine} appear in practice~\cite{buovier2026generative}.
\item \textbf{\emph{Prompt-level transparency}}: Students are asked to give brief explanations and example prompts, encouraging self-reflection 
and offering instructors insight into how GenAI was used.
\item \textbf{\emph{Reflection over punishment}}: the declaration is framed as a description of process, not an admission of guilt, in direct response to empirical evidence that students avoid declaring GenAI use when they perceive the form as a mechanism for penalty rather than transparency~\cite{gonsalves2025addressing}.
\end{itemize}

Next, we describe the two declaration structures and show how they operationalise the original taxonomy in writing and coding assessments.

\subsection{Writing-focused declaration structure}
For writing-focused activities (e.g., a short literature review on technology and society), we designed a declaration form organised around six activity categories, where for each category students (1) state whether GenAI was used (\emph{Yes/No}), (2) specify the extent of use (\emph{Minor, Moderate, Extensive}), and (3)~provide a short explanation with example prompts. The categories are the following:
\begin{itemize}[leftmargin=*,nosep]
\item \textbf{Planning \& Structure}: brainstorming ideas and topics, outlining sections, and structuring arguments.
\item \textbf{Textual Content Generation}: drafting introductions or conclusions, writing body paragraphs, generating examples or case studies.
\item \textbf{Research \& Analysis}: summarising papers, comparing sources, translating non‑English material, clarifying background concepts.
\item \textbf{Revision \& Polish}: grammar and spelling checks, rephrasing sentences, improving clarity and flow, formatting citations and references.
\item \textbf{Visual Content}: creating diagrams or figures, generating charts or tables, suggesting slide layouts for slides.
\item \textbf{Evaluation \& Feedback}: asking GenAI to evaluate argument strength, check logical consistency, or provide feedback on drafts and gaps.
\end{itemize}

These categories operationalise multiple aspects of the Bouvier et al. taxonomy in an essay-writing context~\cite{buovier2026generative}. For example, \emph{Planning \& Structure} and \emph{Visual Content} may involve \emph{Brainstorm} and \emph{Design} aspects; \emph{Textual Content Generation} aligns with \emph{Generate}; \emph{Research \& Analysis} draws on \emph{Interpret} and, at times, \emph{Evaluate}; and \emph{Revision \& Polish} and \emph{Evaluation \& Feedback} relate to \emph{Refine, Get Feedback}, and \emph{Reflect}~\cite{buovier2026generative} Making these distinctions explicit helps foreground the boundary between acceptable assistance (e.g., limited support with revision) and more problematic outsourcing of core content.

\subsection{Coding-focused declaration structure}
For coding assessments, we developed a parallel declaration structure reflecting typical phases of software development and code-based learning. The form includes five activity categories listed below and, 
similarly to the writing-focused structure, students indicate whether GenAI was used and to what extent, providing a short description and example prompts.

\begin{itemize}[leftmargin=*,nosep]
\item \textbf{Planning \& Design}: clarifying problem requirements, sketching system architecture, creating UML diagrams, planning interfaces.
\item \textbf{Code Generation}: generating code skeletons, writing functions or methods, producing unit tests or mock data.
\item \textbf{Code Improvement}: debugging errors, refactoring existing code, optimising performance, requesting code review-style feedback.
\item \textbf{Understanding \& Learning}: explaining error messages, learning syntax or language features, understanding existing codebases, learning algorithms or design patterns.
\item \textbf{Documentation \& Reporting}: writing comments, producing README files, generating user-level documentation or short technical reports.
\end{itemize}

Here the mapping to the taxonomy is tailored to coding practice~\cite{buovier2026generative}. \emph{Planning \& Design} is primarily associated with \emph{Brainstorm} and \emph{Design}; \emph{Code Generation} closely aligns with \emph{Generate}; \emph{Code Improvement} relates to \emph{Refine} and \emph{Get Feedback}; \emph{Understanding \& Learning} foregrounds \emph{Interpret} and \emph{Reflect}; and \emph{Documentation \& Reporting} may involve a combination of \emph{Generate}, \emph{Refine}, and \emph{Interpret}, depending on how GenAI is used~\cite{buovier2026generative}. The structure allows both students and instructors to distinguish between GenAI as a learning aid (e.g., understanding error messages) and GenAI as an authoring tool (e.g., generating core solution code), which carry different implications in an educational setting.

\subsection{Design choices and intended use}
Across both declaration structures, we intentionally retained a consistent pattern: the same extent scale (\emph{Minor, Moderate, Extensive}),  combined with brief explanations and example prompts. The categories in both cases mirror typical stages of work (planning $\rightarrow$ implementation $\rightarrow$ refinement $\rightarrow$ evaluation). The goal at this stage is not to provide a precise measurement instrument, but to offer a structured way for students to describe their GenAI use and for educators to open more nuanced conversations about acceptable assistance, academic integrity, and professional practice. 
The three-point extent scale is defined as follows: \textbf{\emph{Minor}}: GenAI was used once or a small number of times for a limited part of the task; \textbf{\emph{Moderate}}: GenAI was used repeatedly across multiple parts of the task, but substantive decisions and content remained the student's own; \textbf{\emph{Extensive}}: GenAI was integral to completing the task in its submitted form, such that the work would be significantly different without it.

\textbf{Limitations:} The framework relies on accurate self-reporting, which has known limitations. Empirical work has shown that novice programmers can experience an illusion of competence and misattribute GenAI contributions to their own work~\cite{prather2024widening}. In addition, the boundary of what counts as ``GenAI use'' is not always obvious to students: IDE autocomplete, browser-based writing assistants, and search engines with AI-generated summaries all sit on a spectrum with conversational tools like ChatGPT. The declaration should, thus, be accompanied by clear institutional guidance on which tools are in scope. 


\section{Discussion and Implications}
The framework reframes GenAI declarations from a compliance artefact into a reflective instrument. By decomposing a single declaration into activity-specific categories and asking for extent, explanation, and example prompts, it attempts to make the student's workflow visible in a way that a binary checkbox cannot~\cite{gonsalves2025addressing,maguire2025themes}. Three implications are worth highlighting.
 
For \textbf{\emph{students}}, the framework makes the boundary between acceptable assistance and misconduct legible at the point of use, rather than being inferred from general guidance. Categories such as \emph{Textual Content Generation} or \emph{Code Generation} name the activities most likely to raise integrity concerns, while \emph{Revision \& Polish} or \emph{Understanding \& Learning} signal that other uses are legitimately supported. This is intended to reduce the perception, noted in prior work, that any declaration implies admitting wrongdoing \cite{gonsalves2025addressing}.
 
For \textbf{\emph{educators}}, the structure produces disclosures that are directly usable in feedback and course design. A cohort in which most students declare \emph{Extensive} use in \emph{Code Generation} but \emph{Minor} use in \emph{Understanding \& Learning} tells a different pedagogical story from the reverse pattern, and this can inform decisions about scaffolding, assessment redesign, and the level of GenAI use permitted in subsequent tasks~\cite{perkins2024artificial}. The declaration becomes a formative signal as well as a summative record.
 
For \textbf{\emph{institutions}}, the framework is complementary rather than substitutive: it sits below policy-level instruments such as the AIAS, providing the granular disclosure layer that those instruments do not themselves specify~\cite{perkins2024artificial}. 
The same two structures can in principle support very different institutional stances on permitted GenAI use, because the declaration describes practice rather than prescribing it.

The framework also has clear limitations. It relies on accurate self-reporting in a context where students may misattribute GenAI contributions or be uncertain which tools count~\cite{prather2024widening}. It does not itself detect misconduct, and pairing it with detection tools raises well-documented reliability concerns.
And while the categories were derived from an activity taxonomy grounded in computing education~\cite{buovier2026generative}, the transferability to other disciplines remains an open question.

\section{Future Work and Conclusion}
Future work falls into three strands. First, empirical evaluation: comparing student declarations against observed workflow data to assess whether declarations track actual behaviour. Second, pedagogical evaluation: examining whether using the framework changes student reflection, perceived psychological safety around disclosure, and cohort-level patterns of GenAI use over time. Third, cross-disciplinary adaptation: testing whether the underlying principles transfer to disciplines with different assessment patterns, and identifying the minimal set of categories that generalises.
 
In conclusion, binary declarations are an instrument for a workflow that is increasingly multi-stage, multi-tool, and cognitively varied. We have proposed a domain-specific alternative for CS assessment that operationalises an established activity taxonomy~\cite{buovier2026generative} into two parallel, task-specific declaration structures. The contribution is a design artefact and a position: task-specific disclosure is, we argue, a necessary complement to policy-level frameworks if declarations are to support honest assessment, AI literacy, and preparation for professional environments. 

\bibliography{bibfile}

\end{document}